\documentclass[pra,twocolumn,aps]{revtex4}
\usepackage{amsmath}
\usepackage{wrapfig}
\usepackage[pdftex]{graphicx}
\usepackage{wrapfig}
\usepackage[usenames,pdftex]{color}
\usepackage[ansinew]{inputenc}
\usepackage{calrsfs}
\usepackage{amstext}
\usepackage{gensymb}
\definecolor{rem}{rgb}{1.0,0,0}

\begin{document}

\title{Atomic memory based on recoil-induced resonances}

\author{J. C. C. Capella,$^1$ A. M. G. de Melo,$^1$ J.~P.~Lopez,$^{1,2}$ J.~W.~R.~Tabosa,$^1$ and D. Felinto$^1$}

\affiliation{$^1$Departamento de F\'{\i}sica, Universidade Federal de Pernambuco, 50670-901 Recife, PE - Brazil \\ $^2$Departamento de F\'{\i}sica, Universidade Federal da Paraíba, 58059-970 Jo\~ao Pessoa, PB - Brazil}
 

\date{\today}

\begin{abstract}
In this work we perform a detailed theoretical and experimental investigation of an atomic memory based on recoil-induced resonance in cold cesium atoms. We consider the interaction of a  nearly degenerated pump and probe beams with an ensemble of two-level atoms. A full theoretical density matrix calculation in the extended Hilbert space of the internal and external atomic degrees of freedom allows us to obtain, from first principles, the transient and stationary responses determining the probe transmission and the forward four-wave mixing spectra. These two signals are generated together at the same order of perturbation with respect to the intensities of pump and probe beams. Moreover, we have investigated the storage of optical information on the spatial modes of light beams in the atomic external degrees of freedom, which provided a simple interpretation for the previously-reported non-volatile character of this memory. The retrieved signals after storage reveal the equivalent role of probe transmission and four-wave mixing, as the two signals have similar amplitudes. Probe transmission and forward four-wave-mixing spectra were then experimentally measured for both continuous excitation and after storage. The experimental observations are in good agreement with the developed theory and open a new pathway for the reversible exchange of optical information with atomic systems.
\end{abstract}

\maketitle
\section{Introduction} 
\noindent Similarly as light energy and momentum can be transferred to atoms altering their state of motion, the inverse process where the atomic kinetic energy and momentum are transferred to the light field can modify the light field state as well. This last process leads to the observation of various phenomena associated with laser cooled atoms as for instance the so called recoil-induced resonance (RIR) phenomenon, where the exchange of energy and momentum between two light beams is mediated by the atomic external degrees of freedom. This phenomenon was firstly predicted theoretically by Guo et al \cite{Guo92,Guo93} and soon after observed experimentally \cite{Courtois94}. Since then, the RIR phenomenon received considerable attention, and a number of applications associated with it has been demonstrated \cite{WGawlik06}. For instance, RIR was used for temperature diagnostic of cold atomic ensembles both in free atoms at stationary  \cite{Grynberg94,Fischer01}  and transient  \cite{Guibal96} domains, and in atoms confined in optical lattices  \cite{Gawlik06}. More recent applications of RIR for atomic thermometry can be found in Refs.~\cite{YanTing15,Wang15}. It was also used for optical switching~\cite{Gordon10} and to probe the transient dynamic of atoms in 1D optical lattices~\cite{Kozuma95}.

The RIR phenomenon was also employed to observe very high optical gain in anisotropic medium  \cite{Prentiss05}  as well as to observe collective effects as in the collective atomic recoil laser (CARL), which was firstly proposed theoretically in \cite{Bonifacio94} and experimentally demonstrated in \cite{Courteille03}. Using a four-wave mixing (FWM) configuration in a degenerate two-level system of cold cesium atoms, where Zeeman coherence as well as coherence between momentum states via RIR can be excited, our group has observed a giant optical gain and self-oscillation \cite{Lopez19} through coupled  cascading parametric  backward- and forward-FWM (FFWM).

In another context, we have also recently demonstrated the storage of information on the spatial modes of light based on the external atomic degrees of freedom, both using the non-localized degrees of freedom associated with the RIR phenomenon  \cite{Allan16}, as well as the quantized  energy levels of atoms localized in a 1D optical lattice \cite{Lopez17}. This new type of memory using the atomic external degrees of freedom is particularly attractive since it is less sensitive to external magnetic and electric fields. Indeed, differently from the memories based on ground state coherences  associated with the Zeeman sub levels, we have demonstrated its non volatility and robustness to the reading process, which does not destroy the stored information, so its storage time is mainly determined by the atomic motion. Moreover, using the gain mechanism described in \cite{Lopez19}, we have also demonstrated the operation of an atomic memory that can amplify the stored signal during the reading process \cite{Lopez20}. 

In view of these experimental developments in different directions, the aim of the present work is to report a detailed theoretical and experimental investigation of the RIR phenomenon and its role in the storage of light through the modelling and observation of both the probe transmission and FFWM spectra, in the writing and reading phases, considering the simplest case of an ensemble of two-level atoms. Although the RIR effect is sometimes interpreted as Rayleigh scattering into a density grating~\cite{Courtois94}, we followed a different approach that we consider more amenable to model our experimental data. Our first-principles approach models it, in the lowest order of perturbation, as Raman scattering between differently populated momentum states~\cite{Guo92}. This process creates a coherence grating between momentum states of the ground-state manifold, which later scatters the incident optical fields. Most importantly, this process does not require any dislocation of atoms in space to form density gratings, being robust with respect to the power of the excitation fields.

The model and calculation developed here allows us to obtain the transient and stationary responses of the system, and provides a simple interpretation of the non-volatility mentioned above for the RIR memory. We also performed a complete experimental investigation measuring simultaneously the probe transmission and FFWM excitation spectra, as well as the corresponding spectra for the retrieved signals after a specific storage time. These measurements are in good agreement with our theoretical description. This pair of signals constitute an overall quasi-phase matching process that could be explored in the future for its classical and quantum correlations.

The manuscript is organized as follows: in section II  we introduce the basic assumptions of our theoretical model and present the main results of a density matrix calculation performed in the combined internal and external Hilbert-space state. In section III we present our experimental apparatus and measurements; In section IV we compare and discuss the main experimental observations with our theoretical predictions, and finally in section V we present our main conclusions.

\section{Theoretical model} 
\begin{figure}[htb!]  
\includegraphics[scale=0.27]{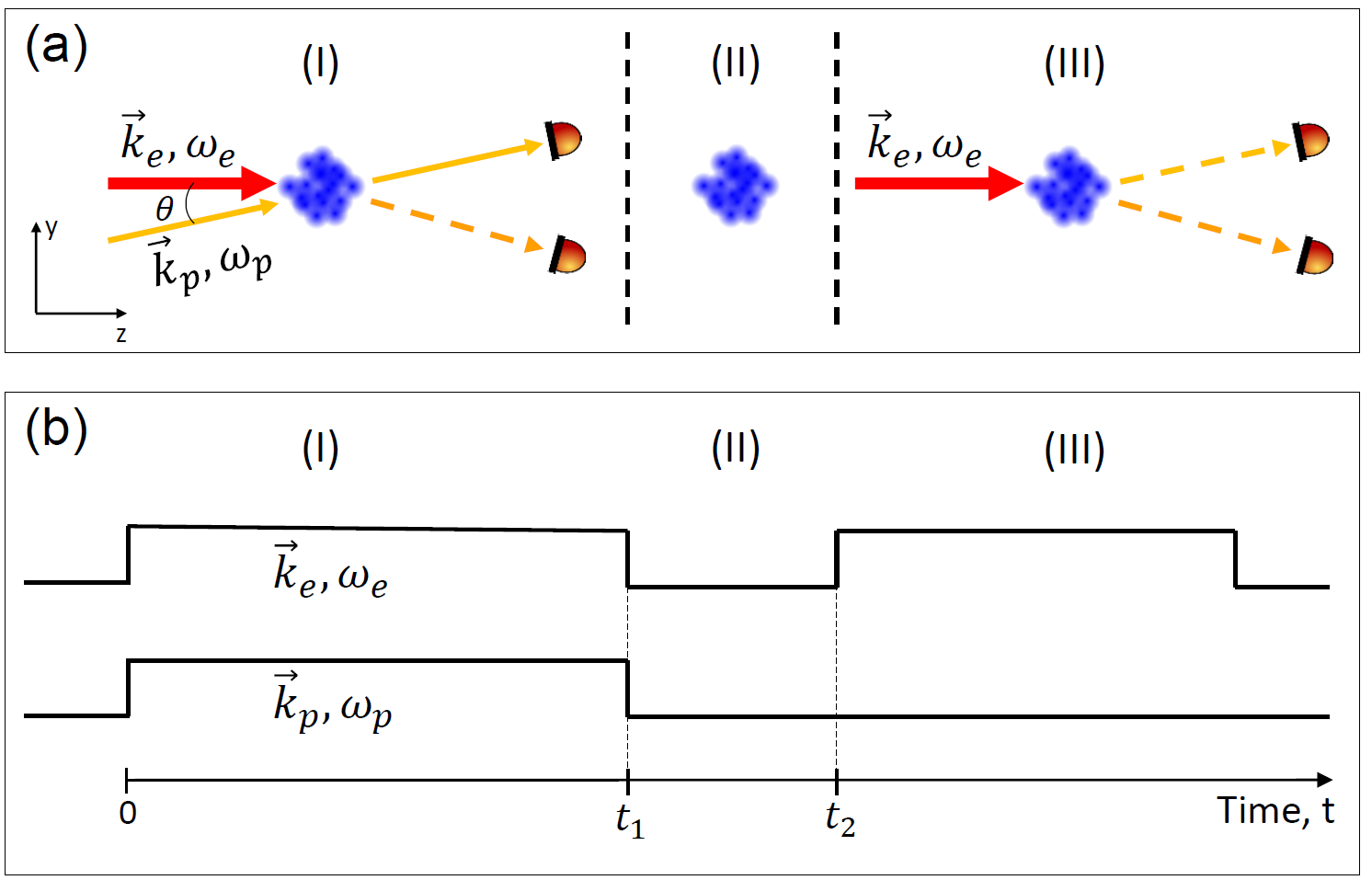}
\caption{(a) Visualization of the three phases of the theoretical model: I) interaction with the excitation $(\vec{k}_e, \omega_e)$ and probe $(\vec{k}_p, \omega_p)$ coplanar fields; II) dark evolution of the system; III) reading phase using only the excitation field. The time duration of the dark phase is upscaled. (b) Excitation time sequence.} \label{fig1}
\end{figure}

In this section we present a theoretical approach to describe the non-volatile robust memory associated with the RIR phenomenon. Explicitly, we consider an initially free cold atomic cloud subjected to three subsequent processes (see Fig.~\ref{fig1}): an interaction with an excitation and probe beams (the writing phase), then a dark evolution where no fields act on the atomic cloud (dark/storage phase), and, finally, an interaction with only the excitation beam (the reading phase). We are interested in the signals generated along the probe beam and the FFWM directions. Further details of the theoretical model for each phase will be given in the respective subsections.

\subsection{The writing phase.}
For the writing phase we consider an ensemble of cold two-level atoms interacting with two fields that can be approximated by plane waves, an (strong) excitation field with wavevector $\vec{k}_e$ and frequency $\omega_e$ and a (weak) probe field with wavevector $\vec{k}_p$ and frequency $\omega_p$:
\begin{equation}
\textbf{E}(t) = \left[{\cal E}_e \cos (\vec{k}_e\cdot \hat{\bf{r}} -\omega_et) +  {\cal E}_p \cos (\vec{k}_p\cdot \hat{\bf{r}} -\omega_pt)\right]\hat{e}\,,
\end{equation}
where ${\cal E}_e$ and ${\cal E}_p$ are the excitation and probe amplitudes, respectively and $\hat{e}$ is the polarization vector common to both fields. The wavevector $\vec{k}_e$ points in the longitudinal $z$ direction, and the $y$ axis will be referred to as the transverse direction. These two define the $zy$ plane containing both excitation and probe fields. We assume a small angle $\theta$ between the directions of the two fields, and excitation and probe frequencies close to the atomic resonance at $\omega_0$, but with detunings $\Delta_e = \omega_0 - \omega_e$ and $\Delta_p = \omega_0 - \omega_p$ much larger than the excited state natural linewidth $\Gamma$, i.e., $\Delta_e,\Delta_p >> \Gamma$. The atomic internal ground state $|1\rangle$ has energy $E_1$, and the internal excited state $|2\rangle$ energy $E_2$. The atom also has a linear momentum $\vec{p}$, associated with a state $|\vec{p}\rangle$ for its external degrees of freedom. 

Considering the ground state energy $E_1=0$, the Hamiltonian for a single atom in the ensemble can be written as:

\begin{equation}
\hat{H} = \frac{\hat{\bf{p}}^{\,2}}{2m} + \hbar\omega_0 |2\rangle \langle 2| - e \,\hat{\bf{D}}\cdot\textbf{E}(t) \,,
\end{equation}
in the dipole interaction approximation, with $\hat{\bf{D}}$ the atomic electric dipole operator and $m$ the atomic mass. Using the basis of the internal states, the Hamiltonian becomes then:

\begin{align}
\hat{H} &= \frac{\hat{\bf{p}}^{\,2}}{2m} + \hbar\omega_0 |2\rangle \langle 2|- \left( \mu_{12} |1\rangle \langle 2| + \mu_{21} |2\rangle \langle 1|\right) \times \nonumber\\
        &\times  \left[ {\cal E}_e \cos (\vec{k}_e\cdot \hat{\bf{r}} -\omega_et) +  {\cal E}_p \cos (\vec{k}_p\cdot \hat{\bf{r}} -\omega_pt) \right] \,, \label{V2}
\end{align}
with $\mu_{12}$ the dipole moment of the $|1\rangle \rightarrow |2\rangle $ transition. Considering both internal and external degrees of freedom, the density matrix for the system is defined as:
\begin{equation}
\hat{\rho} = \sum_{i,j = 1}^2 \int d\vec{p} d\vec{p^{\, \prime}}\,\, \rho_{ij}(\vec{p},\vec{p}^{\,\prime}) |i\rangle |\vec{p}\rangle \langle \vec{p}^{\,\prime}| \langle j| \,,
\end{equation}
with $\rho_{ij}(\vec{p},\vec{p}^{\,\prime})$ providing the general populations and coherences in all degrees of freedom. The density matrix components can be grouped now in five different family of terms:
\begin{enumerate}
\item $\rho_{11}(\vec{p},\vec{p}) \equiv$ ground-state populations; 
\item $\rho_{22}(\vec{p},\vec{p}) \equiv$ excited-state populations;
\item $\rho_{12}(\vec{p},\vec{p}^{\,\prime}) \equiv$ optical coherences;
\item $\rho_{11}(\vec{p},\vec{p}^{\,\prime}) \equiv$ ground-state coherences ($\vec{p} \neq \vec{p}^{\,\prime}$);
\item $\rho_{22}(\vec{p},\vec{p}^{\,\prime}) \equiv$ excited-state coherences ($\vec{p} \neq \vec{p}^{\,\prime}$).
\end{enumerate}
We anticipate that our initial state is restricted to the ground state populations (family 1) obeying a Maxwell-Boltzmann distribution in momentum. The core of the temporal dynamics, however, will unfold in the ground-state coherences (family 4) resulting in the observed traces in the light emitted by the ensemble (from the optical coherences of family 3).

The time-evolution of the density matrix components will be dictated by the Liouville equation:
\begin{equation}
\frac{d \rho_{ij}(\vec{p},\vec{p}^{\,\prime})}{dt} = \frac{i}{\hbar} \langle \vec{p}| \langle i| [\hat{\rho},\hat{H}] |j\rangle |\vec{p}^{\,\prime}\rangle \,.
\end{equation}
Defining a rotating frame where $\rho_{12}(\vec{p},\vec{p}^{\,\prime}) = e^{i\omega_e t}\sigma_{12}(\vec{p},\vec{p}^{\,\prime})$ and using the rotating wave approximation, we obtain:
\begin{widetext}
\begin{align}
\frac{d \rho_{11}(\vec{p},\vec{p}^{\,\prime}, t)}{dt} &= i \Delta(\vec{p}, \vec{p}^{\, \prime}) \rho_{11}(\vec{p},\vec{p}^{\,\prime}, t) -i\left[  \Omega_e\sigma_{12}(\vec{p},\vec{p}^{\,\prime}+\hbar\vec{k}_e, t) +\Omega_p\sigma_{12}(\vec{p},\vec{p}^{\,\prime}+\hbar\vec{k}_p, t)e^{-i\delta t} - \mathrm{h. c.}\right] \,, \label{drho11}\\
\frac{d \sigma_{12}(\vec{p},\vec{p}^{\,\prime}, t)}{dt} &= i\left[\Delta_e + \Delta(\vec{p}, \vec{p}^{\, \prime})\right] \sigma_{12}(\vec{p},\vec{p}^{\,\prime}, t) 
-i  \left\{ \Omega_e\left[\rho_{11}(\vec{p},\vec{p}^{\,\prime}-\hbar\vec{k}_e, t) - \rho_{22}(\vec{p}+\hbar\vec{k}_e,\vec{p}^{\,\prime}, t) \right] + \right.\nonumber\\
&+\left. \Omega_p e^{i\delta t}\left[\rho_{11}(\vec{p},\vec{p}^{\,\prime}-\hbar\vec{k}_p, t) - \rho_{22}(\vec{p}+\hbar\vec{k}_p,\vec{p}^{\,\prime}, t)\right]\right\} \,,
\end{align}
\end{widetext}
where $\Delta(\vec{p}, \vec{p}^{\, \prime}) = \frac{(\vec{p}^{\,\prime 2} - \vec{p}^{\, 2})}{2 m \hbar}$, $\Omega_e = \mu_{12}{\cal E}_e/2\hbar$ and $\Omega_p = \mu_{12}{\cal E}_p/2\hbar$ are the Rabi frequencies associated with the excitation and probe fields, $\delta = \omega_p-\omega_e$ is the two-photon detuning, and $\mathrm{h.c.}$ stands for the hermitian conjugate. We can now express the assumption that the probe field is much weaker than the excitation field as: $\Omega_p << \Omega_e$. Considering a large detuning from the excited state ($\Delta_e >> \Gamma, \delta, p^2/2\hbar m$), we can adiabatically eliminate the excited state, approximating
\begin{align}
\rho_{22}(\vec{p},\vec{p}^{\,\prime}, t) &\approx 0\;, \\
\frac{d \sigma_{12}(\vec{p},\vec{p}^{\,\prime}, t)}{dt} &\approx 0.
\end{align} 
These two conditions result in
\begin{eqnarray}
\sigma_{12}(\vec{p},\vec{p}^{\,\prime}, t) &= \Delta_e^{-1} \left[ \Omega_e\rho_{11}(\vec{p},\vec{p}^{\,\prime}-\hbar\vec{k}_e, t) + \right. \nonumber \\
&+ \left. \Omega_p\rho_{11}(\vec{p},\vec{p}^{\,\prime}-\hbar\vec{k}_p, t)e^{i\delta t} \right]\,. \label{sigma12}
\end{eqnarray}

In this way, we now only need to focus on calculating $\rho_{11}(\vec{p},\vec{p}^{\,\prime},t)$, whose Eq.~\eqref{drho11} becomes:
\begin{align}
\frac{d \rho_{11}(\vec{p},\vec{p}^{\,\prime}, t)}{dt} &= \,i\Delta\left(\vec{p},\vec{p}^{\, \prime}\right)\rho_{11}(\vec{p},\vec{p}^{\,\prime}, t) \,- \nonumber\\
&- i\Omega\left[ \rho_{11}(\vec{p},\vec{p}^{\,\prime}+\hbar\Delta\vec{k}, t)e^{i\delta t} + \nonumber \right. \\
&+ \left.\rho_{11}(\vec{p},\vec{p}^{\,\prime}-\hbar\Delta\vec{k}, t)e^{-i\delta t}  - \mathrm{h.c.} \right]\, ,\label{drho11_b}
\end{align}
with $\Delta \vec{k} = \vec{k}_e-\vec{k}_p$ and $\Omega = \Omega_e\Omega_p/\Delta_e$. Equation~\eqref{drho11_b} generates the whole dynamics between different atomic momentum states. It is important to note that we have completely eliminated the dynamics of the internal degrees of freedom of the atom, making it explicit that the RIR phenomenon arises primarily from the dynamics of the external degrees of freedom. For perturbative calculations, a more suitable form of Eq.~\eqref{drho11_b} is
\begin{align}\label{eq14}
\rho_{11}(\vec{p},\vec{p}^{\,\prime},t) &= \rho_{11}(\vec{p},\vec{p}^{\,\prime},0) + \nonumber \\
&+ i \Omega \left\{\int_{0}^{t} d t^{\, \prime} G_{0}(t,t^{\, \prime})\left[ \rho_{11}(\vec{p},\vec{p}^{\,\prime}+\hbar\Delta\vec{k},t^{\, \prime})e^{i\delta t^{\, \prime}} + \right. \right. \nonumber \\
&+\left. \left. \rho_{11}(\vec{p},\vec{p}^{\,\prime}-\hbar\Delta\vec{k},t^{\, \prime})e^{-i\delta t^{\, \prime}} - \mathrm{h. c.}\right]\right\} 
\end{align}
where $G_{0}(t,t^{\, \prime}) = e^{i \Delta(\vec{p},\vec{p}^{\,\prime}) t} e^{- i \Delta(\vec{p},\vec{p}^{\,\prime}) t^{\prime}}$ is a Green's function for the operator $\left[\frac{d}{dt} - i\Delta\left(\vec{p},\vec{p}^{\, \prime}\right)\right]$, and $\Omega$ becomes the perturbation parameter.

We are interested in a first-order solution for $\rho_{11}(\vec{p},\vec{p}^{\,\prime},t)$, which means we should substitute the $0^{\mathrm{th}}$-order solution $\rho^{0}_{11}(\vec{p},\vec{p}^{\,\prime},t) = \rho_{11}(\vec{p},\vec{p}^{\,\prime},0)$ in the integrand. At $t=0$, however, we have no coherences stablished between different momenta states of the atom, only populations. This condition can be expressed as:
\begin{equation}
\rho_{11}(\vec{p},\vec{p}^{\,\prime},0) = \rho_{11}^{0}(\vec{p})\delta(\vec{p} - \vec{p}^{\,\prime}),
\end{equation}
where $\rho_{11}^{0}(\vec{p})$ is given by the standard Maxwell-Boltzmann distribution for the momenta of the atomic ensemble: 
\begin{equation}\label{maxwell}
\rho_{11}^0(\vec{p}) = \frac{m}{\sqrt{(2\pi)^3} \,p_u} e^{-\vec{p}^{\, 2}/2p_u^2} \,,
\end{equation}
with $p_u = \sqrt{m k_BT}$ and $T$ the ensemble's temperature. We obtain then: 
\begin{widetext}
\begin{align} \label{rho11def}
\rho_{11}(\vec{p},\vec{p}^{\,\prime},t) &= \rho_{11}^{0}(\vec{p})\delta(\vec{p} - \vec{p}^{\,\prime}) - \Omega \left\{ \frac{1}{-\delta + \Delta(\vec{p},\vec{p}^{\,\prime})}\delta(\vec{p} - \vec{p}^{\,\prime} - \hbar \Delta\vec{k})\left[\left(e^{i \delta t} - e^{i \Delta(\vec{p},\vec{p}^{\,\prime}) t}\right)\times\left(\rho_{11}^{0}(\vec{p}) - \rho_{11}^{0}(\vec{p} - \hbar\Delta\vec{k})\right)\right]\right. \nonumber +\\
                                         &+ \left.  \frac{1}{\delta + \Delta(\vec{p},\vec{p}^{\,\prime})}\delta(\vec{p} - \vec{p}^{\,\prime} + \hbar \Delta\vec{k})\left[\left(e^{-i \delta t} - e^{i \Delta(\vec{p},\vec{p}^{\,\prime}) t}\right)\times\left(\rho_{11}^{0}(\vec{p}) - \rho_{11}^{0}(\vec{p} + \hbar\Delta\vec{k})\right)\right]\right\},
\end{align}
\end{widetext}
 
which fully determines $\rho_{11}(\vec{p},\vec{p}^{\,\prime},t)$ in first-order approximation. Note that $\rho_{11}(\vec{p},\vec{p}^{\,\prime},t)$ represents a ground-state population for the internal degrees of freedom of the atoms, but describes both populations and coherences of the momentum states describing the atomic external degrees of freedom. These terms act as ``sources" for the optical coherences as in Eq. \eqref{sigma12}. In this order of perturbation, they take into account effects that depend on one single ``kick" of momentum, $\hbar \Delta\vec{k}$, generating signals in different directions. If we were interested in a second-order solution for $\rho_{11}(\vec{p},\vec{p}^{\,\prime},t)$, we would proceed in an analogous manner, substituting the first order solution we found onto the integrand of \eqref{eq14}. This would lead to second-order terms taking into account now two kicks of momentum (and a correction for the ``zero-kicks'' solution) that would result, in the conditions considered here, in the creation of much weaker signals in other directions. 

After substitution of \eqref{rho11def} into \eqref{sigma12}, we now obtain the usual local optical coherence, given by \cite{Guo92}:
\begin{equation}
\rho_{12}(\vec{r};t) = \frac{1}{(2\pi\hbar)^3} \int d\vec{p}d\vec{p}^{\,\prime}e^{i\frac{\vec{p}\cdot\vec{r}}{\hbar}}\rho_{12}(\vec{p},\vec{p}^{\,\prime},t)e^{-i\frac{\vec{p}^{\,\prime}\cdot\vec{r}}{\hbar}} \,.
\end{equation}  
Since $\Omega_p << \Omega_e$, we disregard terms of order $\Omega \Omega_p$, yielding:
\begin{widetext}
\begin{align}\label{localcoherence}
\rho_{12}(\vec{r};t) &=  \frac{1}{\Delta_{e} (2 \pi \hbar)^{\frac{3}{2}}} \int d\vec{p} d\vec{p}^{\, \prime} e^{\frac{i}{\hbar}(\vec{p} - \vec{p}^{\, \prime})\cdot \vec{r}}\Bigg\{\Omega_{p} e^{i \omega_{p}t} \rho_{11}^{0}(\vec{p})\delta(\vec{p} - \vec{p}^{\,\prime} + \hbar \vec{k}_{p}) + \Omega_{e} e^{i \omega_{e}t}\rho_{11}^{0}(\vec{p})\delta(\vec{p} - \vec{p}^{\,\prime} + \hbar \vec{k}_{e})\, -  \nonumber \\
&- \Omega\Omega_{e} e^{i \omega_{e}t} \Bigg[ \frac{1}{-\delta + \Delta(\vec{p},\vec{p}^{\,\prime} - \hbar \vec{k}_{e})}\delta(\vec{p} - \vec{p}^{\,\prime} - \hbar \Delta\vec{k} + \hbar \vec{k}_{e})  \left(e^{i \delta t} - e^{i \Delta(\vec{p},\vec{p}^{\,\prime} - \hbar \vec{k}_{e}) t}\right)\left(\rho_{11}^{0}(\vec{p}) - \rho_{11}^{0}(\vec{p} - \hbar\Delta\vec{k})\right) + \nonumber \\
                     &+\frac{1}{\delta + \Delta(\vec{p},\vec{p}^{\,\prime} - \hbar \vec{k}_{e})}\delta(\vec{p} - \vec{p}^{\,\prime} + \hbar \Delta\vec{k} + \hbar \vec{k}_{e}) \left(e^{-i \delta t} - e^{i \Delta(\vec{p},\vec{p}^{\,\prime} - \hbar \vec{k}_{e}) t}\right)\left(\rho_{11}^{0}(\vec{p}) - \rho_{11}^{0}(\vec{p} + \hbar\Delta\vec{k})\right) \Bigg]\Bigg\} \,.                  
\end{align}
\end{widetext}
All observed signals originate from this equation, which deserves a closer look. The Dirac deltas express the momentum conservation required for any process. The first two terms give the linear response of the atomic medium, while the last two terms originate from the $3^{\mathrm{rd}}$-order non-linearity of the medium in momentum space. We turn our attention to the latter noting that the first term inside the brackets describes the gain-attenuation of the probe field, since $- \hbar \Delta\vec{k} + \hbar \vec{k}_{e} = \hbar \vec{k}_{p}$. The second term inside the brackets, on the other hand, describes a FFWM generation process. We can see that the efficiency of these processes are proportional to the difference in population of the external levels $|\vec{p}\rangle$ and $|\vec{p} \pm \hbar \Delta \vec{k}\rangle$. Moreover, note that these two processes arise at the same order of perturbation which goes to show that a more complete understanding of the RIR phenomenon should consider both processes as equivalent.
\vspace{-0.6cm}
\subsubsection{Transmission signal.}
To obtain the transmission signal, we need to look at the plane waves seeded in the $\vec{k}_{p}$ direction. Those arise from the terms containing $\delta(\vec{p} - \vec{p}^{\,\prime} - \hbar \Delta \vec{k} + \hbar \vec{k}_e) = \delta(\vec{p} - \vec{p}^{\,\prime} + \hbar \vec{k}_{p})$ in \eqref{localcoherence}. Explicitly, we have:
\begin{align}
&\rho_{12}^{p}(\vec{r};t) = \frac{\Omega \Omega_{e} e^{-(i\vec{k}_{p}\cdot \vec{r} - \omega_{p}t)}}{\Delta_{e} (2 \pi \hbar)^{\frac{3}{2}}} \int d\vec{p} \,\frac{1}{-\delta + \Delta(\vec{p}, \vec{p} - \hbar \Delta\vec{k})}  \nonumber \\
&\times \left(1 - e^{i (\Delta(\vec{p},\vec{p} - \hbar \Delta\vec{k}) - \delta) t}\right) \left(\rho_{11}^{0}(\vec{p}) - \rho_{11}^{0}(\vec{p} - \hbar \Delta\vec{k})\right). 
\end{align}
Note that using the small angle approximation, the momentum exchange only happens on the transverse $y$ direction which implies that:
\begin{equation}\label{approx1}
\Delta(\vec{p}, \vec{p} - \hbar \Delta\vec{k}) = \frac{\left(|\vec{p}-\hbar\Delta\vec{k}|^2 -p^2\right)}{2 m \hbar} \approx -\frac{p_y\Delta k}{m} + \frac{\hbar}{2 m}\Delta k^{2} \,.
\end{equation}
Now, if such momentum exchange is small when compared to the average momentum of atoms, we may also approximate:
\begin{align}\label{approxmaxwell}
\rho_{11}^{0}(p_y) - \rho_{11}^{0}(p_y - \hbar \Delta k) &\simeq (- \hbar \Delta k) \cdot \frac{\partial \rho_{11}^0(p_y)}{\partial p_y},
\end{align}
where we use the $1\mathrm{D}$ version of \eqref{maxwell}. The probe transmission signal is proportional to the imaginary part of the  slowly-varying coherence, $\sigma_{12}^p(\vec{r},t) = e^{i\vec{k}_{p}\cdot \vec{r} + i \omega_p t} \rho_{12}^{p}(\vec{r};t)$. Using \eqref{approx1} and \eqref{approxmaxwell}, we obtain then:
\begin{align}
&{\rm Im}\left[ \sigma_{12}^p(\vec{r},t) \right] = -\frac{\Omega \Omega_e ( \, m \Delta k)t}{\Delta_e (2 \pi)^2 \hbar^{\frac{1}{2}} p_{u}^{2}} \times \nonumber \\
&\times \int dp_{y} \, \mathrm{sinc}{[(-\delta -p_{y}\frac{\Delta k}{m} + \frac{\hbar}{2 m}\Delta k^{2}) t]} e^{-\frac{p_{y}^{2}}{2 p_{u}^2}} p_{y}\,.
\end{align}
This expression for the transmission signal has already been found in the literature \cite{Guibal96} where the term of second order in $\Delta k$ was discarded.
\begin{figure}[htb]
\includegraphics[scale=0.3]{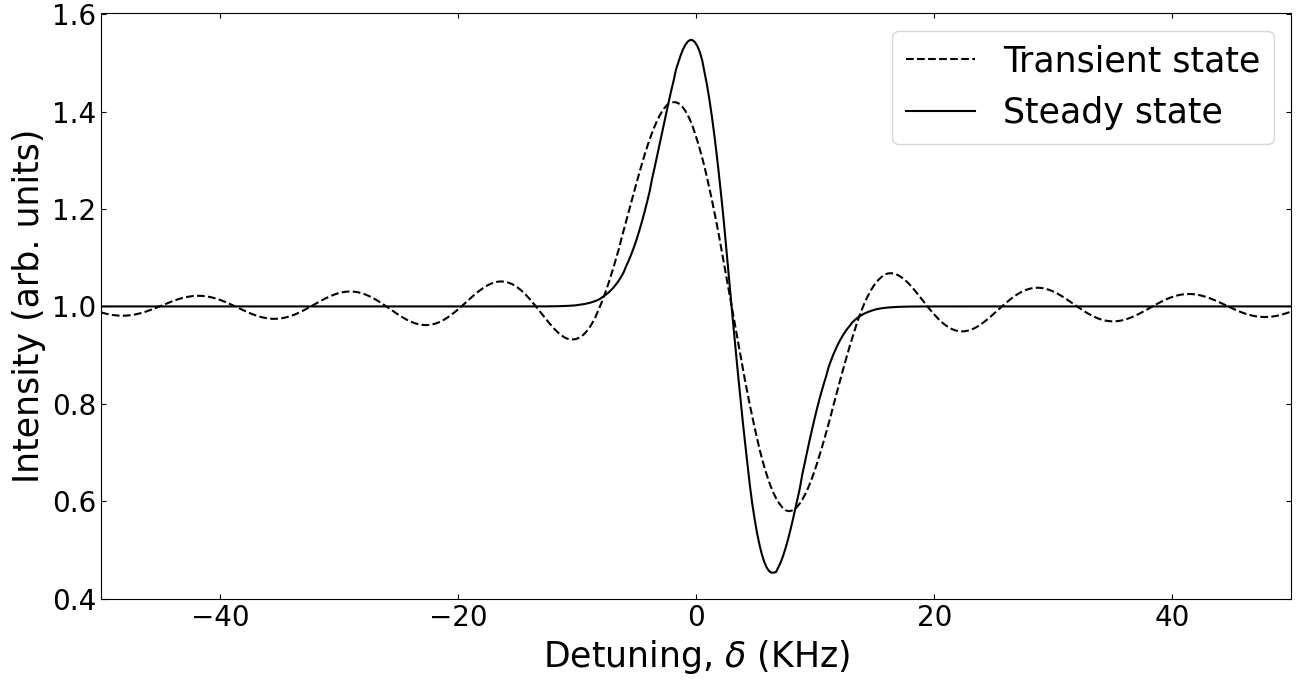}
\caption{Theoretical prediction for the transmission spectrum at $t\simeq 100$ $\mu$s (dashed line) and at $t \gg \tau $ (solid line).} \label{fig2}
\end{figure}
Figure~\ref{fig2} shows the transient and stationary transmission spectra predicted by the theory using $\Delta k = 1.2 \times 10^5 \, \mathrm{m^{-1}}$, a temperature of $\mathrm{T} = 500 \, \mathrm{\mu K}$, and \textit{m} as the Cesium mass.  $\Delta k$ was calculated considering an angle $\theta = 1^{\circ}$ between excitation and probe fields. We will use this set of parameters throughout all theoretical considerations. Note that the other parameters only amount to a global scaling factor, which is irrelevant to the discussion here presented. To obtain a rough estimate of the time $\tau$ it takes for the system to achieve a stationary state, we take the inverse of the corresponding Doppler width, to find $\tau \approx 325 \, \mathrm{\mu s}$. This Doppler width is then directly responsible for the width of the stationary spectrum. Note that the RIR spectrum in Fig.~\ref{fig2} is broader for $t < \tau$, since it is then limited by the interaction time window itself. We do not include any phenomenological homogeneous decay in our model, meaning that the stationary state is reached purely due to the inhomogeneous dephasing of the various atomic velocity groups. In order to highlight this property, we plot in Fig.~\ref{fig3} the time evolution of the transmission signal of a single velocity group (with $p_y = p_u$) versus the evolution of the ensemble of velocity groups, for a single detuning $\delta$. The signal for a single velocity group should describe then a simple oscillation, while the signal for the ensemble reaches the usually observed stationary state.
\begin{figure}[htb]
\includegraphics[scale=0.3]{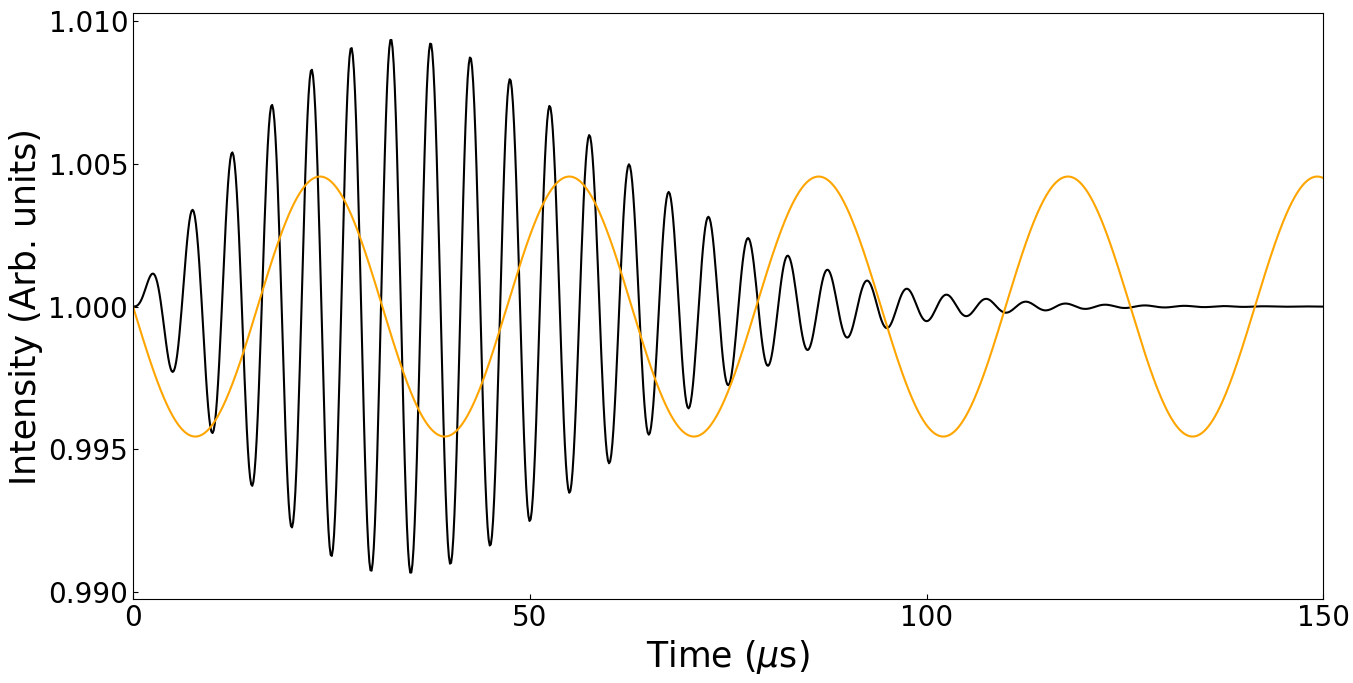}
\caption{(Color online) Transmission signal (black) and singled-out momentum component (upscaled) $p_y = p_u$, in orange/light gray. We use a detuning of $\delta = 200\, \mathrm{kHz}$.} \label{fig3}
\end{figure}

\subsubsection{Forward four-wave mixing.}
Now, we can turn to the problem of obtaining the FFWM signal. To do so, we need to look at the contributions in the $2 \vec{k}_{e} - \vec{k}_p$ direction. Those arise from the terms containing $\delta(\vec{p} - \vec{p}^{\,\prime} + \hbar \Delta \vec{k} + \hbar \vec{k}_e) = \delta(\vec{p} - \vec{p}^{\,\prime} + \hbar (2 \vec{k}_{e} - \vec{k}_p))$. Explicitly, we have: 

\begin{align}
&\rho_{12}^{FWM}(\vec{r},t) = \frac{\Omega \Omega_e (\hbar \, m \Delta k)}{\Delta_e (2 \pi \hbar)^{\frac{3}{2}} p_{u}^{2}}e^{i \omega_{e} t} e^{-i (2 \vec{k}_{e} - \vec{k}_{p})\cdot \vec{r}} e^{-i \delta t}  \nonumber \\
&\times \int dp_{y} \,  \left(1 - e^{i (\delta + p_{y} \frac{\Delta k}{m} + \hbar \frac{\Delta k^2 }{2m})t}\right) \, \frac{e^{-\frac{p_{y}^{2}}{2 p_{u}^2}} p_{y}}{\delta + p_{y} \frac{\Delta k}{m} + \hbar \frac{\Delta k^2 }{2m}} \,.
\end{align}
The FFWM signal is then proportional to the modulus squared of $\rho_{12}^{FWM}(\vec{r},t)$, yielding the spectrum presented in Fig. $4$. As we can see, the theory generates a symmetric spectrum with a peak at $\delta \approx 0$, which evolves in time. As with the structure in Fig.~\ref{fig2}, the central peak becomes narrower as it approaches the stationary state. Figure $4$ also presents, in its inset, theoretical predictions for the FFWM signal evolution in time. These explicitly show the constructive interference resulting in the maximum of the central peak, and the oscillation coming from partial interference defining the values on the side of the peak.

\begin{figure}[htb]
\includegraphics[scale=0.3]{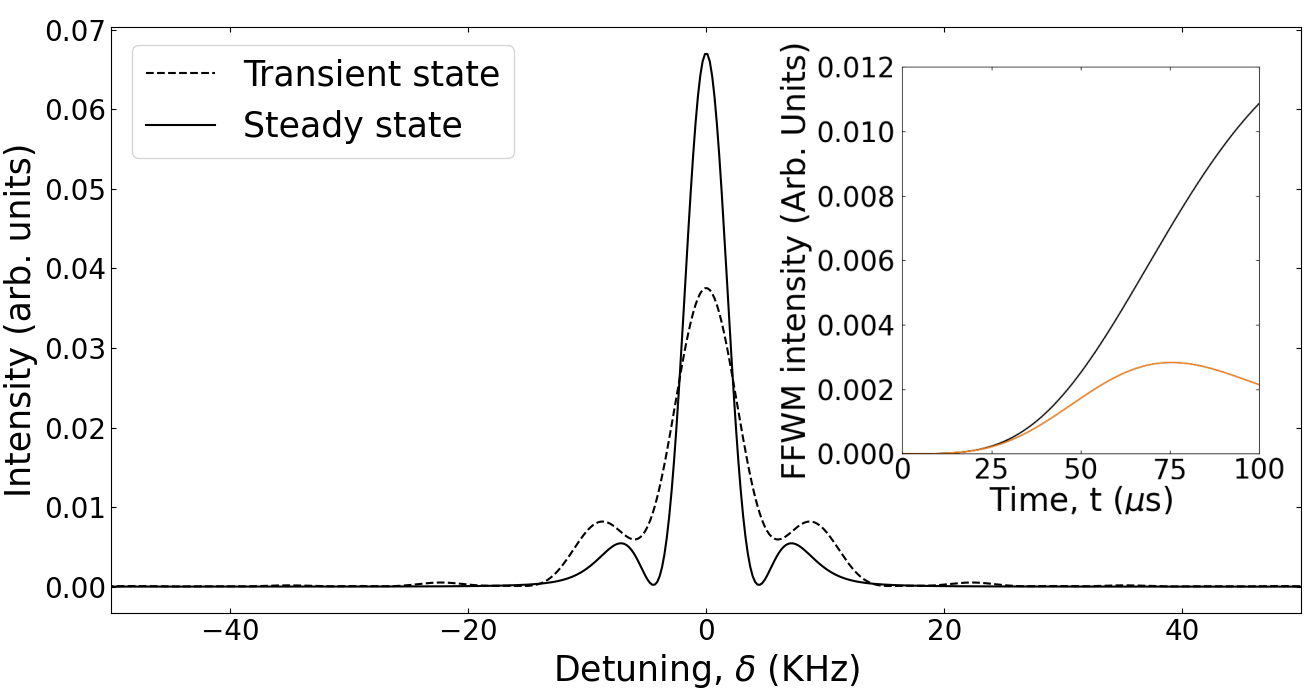}
\caption{(Color online) Theoretical prediction for the FFWM spectrum at $t\simeq 100$ $\mu$s (dashed line) and at $t \gg \tau$ (solid line). Inset: FFWM temporal evolution for $\delta = 0$ (black) and $\delta = \pm 8 \, \mathrm{kHz}$ (orange/light gray).}
\end{figure}

The spectra shown here give the global behavior of the FFWM and transmission signals when continuously generated by the excitation and probe fields. A further study of how these spectra evolve in time will be given in Sec.~\ref{discussions}. In practice, even though it is simpler to calculate and interpret the stationary states of the signals described above, it is quite common to be restricted to transient spectra in cold atoms. As the ensemble becomes colder, the times to reach the stationary state becomes longer and may become on the order of or even surpass typical times for spurious optical pumping to any dark state in the system. This will be the case for the experiments described below.

\subsection{The dark phase.}
We now turn to the modeling of the dark phase, where no fields act on the atomic medium. From now on, we shall use $\rho_{ij}^{\mathrm{I}}(\vec{p},\vec{p}^{\, \prime}, t)$ for coherences and populations in the first (writing) phase of the theoretical model, $\rho_{ij}^{\mathrm{II}}(\vec{p},\vec{p}^{\, \prime}, t)$ for the second (dark) phase and $\rho_{ij}^{\mathrm{III}}(\vec{p},\vec{p}^{\, \prime}, t)$ for the third (reading) phase. Turning back to \eqref{drho11_b} and \eqref{sigma12}, and noting that turning off both fields means $\Omega_e = \Omega_p = \Omega = 0$, these equations simplify to:
\begin{align}
\frac{d\rho^{\rm{II}}_{11}(\vec{p},\vec{p}^{\, \prime},t)}{dt} &= i \Delta(\vec{p}, \vec{p}^{\, \prime}) \,\rho^{\rm{II}}_{11}(\vec{p},\vec{p}^{\, \prime},t)  \nonumber \, ,\\
\sigma^{\rm{II}}_{12}(\vec{p},\vec{p}^{\, \prime},t) &= 0 \,,
\end{align}
whose solution is:
\begin{equation}\label{rho112}
\rho^{\rm{II}}_{11}(\vec{p},\vec{p}^{\, \prime},t)  = \rho^{\rm{I}}_{11}(\vec{p},\vec{p}^{\, \prime},t_1) \, e^{i \Delta(\vec{p}, \vec{p}^{\, \prime})(t - t_1)} \, ,
\end{equation}
where we take $t = t_1$ as the instant the fields are turned off, and assume continuity of the solution throughout the process. We also get directly,
\begin{equation}
\rho^{\rm{II}}_{12}(\vec{p},\vec{p}^{\, \prime},t)  = 0 \,,
\end{equation}
which immediately tell us that we should expect no signals at all to be observed, as there is no optical polarisation. Although we may not be able to promptly measure it, equation \eqref{rho112} shows that in fact the information that was written in the first phase of the experiment is stored in the atomic medium when we turn off the writing fields in the form of coherences between different momentum states. All that is left to do now is the retrieval of this information.

Note that the model predicts that the reading results depend on the storage time because, in the dark period, the off-diagonal elements of the density matrix in the ground-state manifold all evolve with different frequencies, given by the various $\Delta (\vec{p},\vec{p}^{\,\prime})$ when $\vec{p} \neq \vec{p}^{\,\prime}$. Thus the evolution in the dark phase can be understood exactly as a free-induction decay, with $\Delta (\vec{p},\vec{p}^{\,\prime})$ providing the free-evolution phases of the different energy states of the inhomogeneous system at finite temperature.

\subsection{The reading phase.}
In the reading phase, only the excitation field is turned on while the probe field stays off. This implies $\Omega_e \neq 0$ and $\Omega_p = \Omega = 0$. Equations \eqref{drho11_b} and \eqref{sigma12} now imply in: 

\begin{align}
\frac{d\rho^{\rm{III}}_{11}(\vec{p},\vec{p}^{\, \prime},t)}{dt} &= i \Delta(\vec{p}, \vec{p}^{\, \prime}) \,\rho^{\rm{III}}_{11}(\vec{p},\vec{p}^{\, \prime},t) \nonumber \,, \\
\sigma^{\rm{III}}_{12}(\vec{p},\vec{p}^{\,\prime}, t) &= \frac{\Omega_e}{\Delta_e}\, \rho^{\rm{III}}_{11}(\vec{p},\vec{p}^{\,\prime}-\hbar\vec{k}_e, t)\,. \label{sigma12reading}
\end{align}
We already note that the time evolution for $\rho^{\rm{III}}_{11}(\vec{p},\vec{p}^{\, \prime},t)$ does not change from the dark phase, that is, our coherences/populations seem to be unaffected by the reading process. This is the first sign of the robustness of the RIR based memory modeled here. We can solve for $\rho^{\rm{III}}_{11}(\vec{p},\vec{p}^{\, \prime},t)$ directly, obtaining:
\begin{equation}
\rho^{\rm{III}}_{11}(\vec{p},\vec{p}^{\, \prime},t) = \rho^{\rm{II}}_{11}(\vec{p},\vec{p}^{\, \prime},t_2)e^{i \Delta(\vec{p}, \vec{p}^{\, \prime}) (t - t_2)} \,.
\end{equation}
But, note that:
\begin{equation}
\rho^{\rm{II}}_{11}(\vec{p},\vec{p}^{\, \prime},t_2) = \rho^{\rm{I}}_{11}(\vec{p},\vec{p}^{\, \prime},t_1) e^{i \Delta(\vec{p},\vec{p}^{\, \prime})(t_2 - t_1)} \,,
\end{equation}
\\ which then implies that:
\begin{align}\label{rho11III}
\rho^{\rm{III}}_{11}(\vec{p},\vec{p}^{\, \prime},t) &= \rho^{\rm{II}}_{11}(\vec{p},\vec{p}^{\, \prime},t_2)e^{i \Delta(\vec{p}, \vec{p}^{\, \prime}) (t - t_2)} \nonumber \\
&= \rho^{\rm{I}}_{11}(\vec{p},\vec{p}^{\, \prime},t_1) e^{i \Delta(\vec{p},\vec{p}^{\, \prime})(t_2 - t_1)} e^{i \Delta(\vec{p}, \vec{p}^{\, \prime}) (t - t_2)} \nonumber \\
&= \rho^{\rm{I}}_{11}(\vec{p},\vec{p}^{\, \prime},t_1) e^{i \Delta(\vec{p},\vec{p}^{\, \prime})(t - t_1)} \,.
\end{align}
And consequently, 
\begin{equation}\label{memorycoherence}
\sigma^{\rm{III}}_{12}(\vec{p},\vec{p}^{\,\prime}, t) = \frac{\Omega_e}{\Delta_e}\, \rho^{\rm{I}}_{11}(\vec{p},\vec{p}^{\, \prime} - \hbar\vec{k}_e,t_1) e^{i \Delta(\vec{p},\vec{p}^{\, \prime} - \hbar\vec{k}_e)(t - t_1)} \,,
\end{equation}
for $t \geq t_2$. This is the main result of this section: our theoretical model shows in a very direct manner that the reading process does not destroy any information stored in the atomic medium and the retrieved signal is completely indifferent to the moment $t = t_2$ at which the reading process begins and depends only on the time frame of the writing phase, $t_1$. In this way, we provide a theoretical picture that explains the core experimental observations reported in \cite{Allan16}. We now proceed to take a closer look at each mode generated from the atomic memory.

\subsubsection{Retrieved transmission signal.}
Using eq. \eqref{rho11def}, we may write:

\begin{align}
&\sigma^{\rm{III}}_{12}(\vec{p},\vec{p}^{\,\prime}, t) = \frac{\Omega_e}{\Delta_e}\, e^{i \Delta(\vec{p},\vec{p}^{\, \prime} - \hbar\vec{k}_e)(t - t_1)} \left\{\rho_{11}^{0}(\vec{p})\delta(\vec{p} - \vec{p}^{\,\prime} + \hbar\vec{k}_e)\right. \nonumber \\
&- \Omega \left[\rho^{\rm{I (\mathrm{p})}}_{11}(\vec{p},\vec{p}^{\, \prime} - \hbar\vec{k}_e,t_1) \right. \delta(\vec{p} - \vec{p}^{\,\prime} - \hbar \Delta\vec{k} + \hbar\vec{k}_e) - \nonumber \\
&-  \rho^{\rm{I (\mathrm{FWM})}}_{11}(\vec{p},\vec{p}^{\, \prime} - \hbar\vec{k}_e,t_1) \left. \left.\delta(\vec{p} - \vec{p}^{\,\prime} + \hbar \Delta\vec{k} + \hbar\vec{k}_e) \right]\right\},
\end{align}
where we defined:

\begin{align}
&\rho^{\rm{I (\mathrm{p})}}_{11}(\vec{p},\vec{p}^{\, \prime},t_1) = \frac{1}{-\delta + \Delta(\vec{p},\vec{p}^{\,\prime})} \nonumber \\
&\times\left(e^{i \delta t_1} - e^{i \Delta(\vec{p},\vec{p}^{\,\prime}) t_1}\right) \left(\rho_{11}^{0}(\vec{p}) - \rho_{11}^{0}(\vec{p} - \hbar\Delta\vec{k})\right) \nonumber \, ,\\ \\
&\rho^{\rm{I (\mathrm{FWM})}}_{11}(\vec{p},\vec{p}^{\, \prime},t_1) = \frac{1}{\delta + \Delta(\vec{p},\vec{p}^{\,\prime})} \nonumber \\
&\times\left(e^{-i \delta t_1} - e^{i \Delta(\vec{p},\vec{p}^{\,\prime}) t_1}\right) \left(\rho_{11}^{0}(\vec{p}) - \rho_{11}^{0}(\vec{p} + \hbar\Delta\vec{k})\right) \,.
\end{align}
These will generate, respectively, the retrieved signal in the direction of the probe field and in the FFWM direction. The component in the probe direction of the local optical coherence is then given by:
\begin{align}
\rho^{\rm{III(\mathrm{p})}}_{12}(\vec{r}, t) &= -\frac{\Omega \Omega_e}{\Delta_e(2 \pi \hbar)^{\frac{3}{2}}}e^{i \omega_e t}e^{-i \vec{k}_p \cdot \vec{r}} \nonumber \\
&\int d\vec{p}\,e^{i \Delta(\vec{p},\vec{p} - \hbar\Delta\vec{k})(t - t_1)}\rho^{\rm{I (\mathrm{p})}}_{11}(\vec{p},\vec{p} - \hbar\Delta\vec{k},t_1) \,,
\end{align}
With the measured signal being proportional to the modulus squared of $\rho^{\rm{III(\mathrm{p})}}_{12}(\vec{r}, t)$. The blue curves in Fig.~\ref{fig:memory} presents the theoretical spectra (panel a) and temporal evolutions (panel b) of the retrieved probe signal. Note that the spectra of the retrieved probe beam is now a peak, similarly to the continuously-generated FFWM spectra, as they are generated from similarly excited coherences between external momentum states of the atoms. Moreover, the temporal evolution reveals a decay much shorter than the time to reach the stationary state in the writing process for these same conditions (Fig.~\ref{fig2}). The time of tens of microseconds for the decay is related to the diffusion time of atoms between fringes of the coherence grating printed in the ensemble, as already pointed out in~\cite{Allan16}.

\subsubsection{Retrieved FFWM signal.}
\begin{figure}[!htb]
	\includegraphics[scale=0.3]{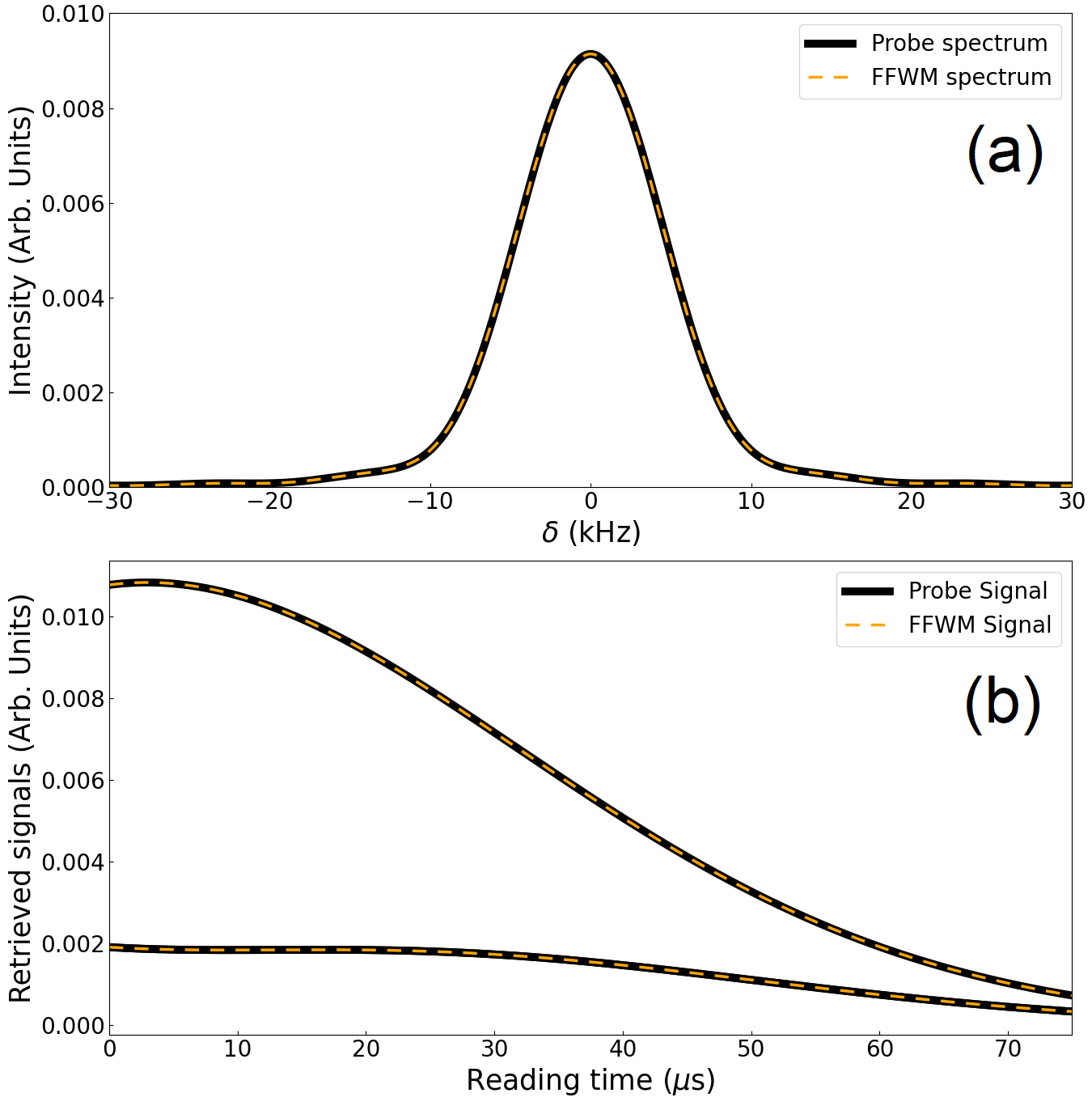}
	\caption{(a) Theoretical spectra for the signals in the probe and FFWM direction, taken at $t = 133\, \mathrm{\mu s}$ considering $t_1 = 102 \, \mathrm{\mu s}$ and $t_2 = 107 \, \mathrm{\mu s}$, which corresponds to a reading time of about $26 \, \mathrm{\mu s}$. (b) Theoretical time profiles for the signals in the probe and FFWM direction for (from top to bottom): $\delta \approx 0$ and $\delta \approx \pm 8 \, \mathrm{kHz}$. Note that the last two are superimposed due to the symmetric nature of the spectrum.}
	\label{fig:memory}
\end{figure}

Now, it is straightforward to obtain the retrieved FFWM signal. In fact, we may write the FFWM-component of the local optical coherence as:
\begin{align}
&\rho^{\rm{III(\mathrm{FWM})}}_{12}(\vec{r}, t) = -\frac{\Omega \Omega_e}{\Delta_e(2 \pi \hbar)^{\frac{3}{2}}}e^{i \omega_e t}e^{-i (2 \vec{k}_e -\vec{k}_p) \cdot \vec{r}} \nonumber \\
&\int d\vec{p}\,e^{i \Delta(\vec{p},\vec{p} + \hbar\Delta\vec{k})(t - t_1)} \rho^{\rm{I (\mathrm{FWM})}}_{11}(\vec{p},\vec{p} + \hbar\Delta\vec{k},t_1) \,.
\end{align}
The measured signal is then also given by the modulus squared of $\rho^{\rm{III(\mathrm{FWM})}}_{12}(\vec{r}, t)$, which generates the spectrum shown in fig. \ref{fig:memory}. Note the peak at $\delta \approx 0$ and it's symmetric structure, just as the spectrum in the probe direction. We may also investigate the time profile of this signal and we note again the similarities between the signals in the FFWM and probe directions. As we pointed out above, this is no coincidence: our theory shows that these signals originate from the same process and have the same behavior. The only differences we saw in the writing phase were due to the manner in which we observed each signal. In the reading phase, the measurements are performed the same way in both directions, and the differences we once observed vanish. Figure \ref{fig:memory} explicits this behavior.

\section{Experiment and Results} 

\begin{figure}[!htb]
	\fbox{\includegraphics[scale=0.35]{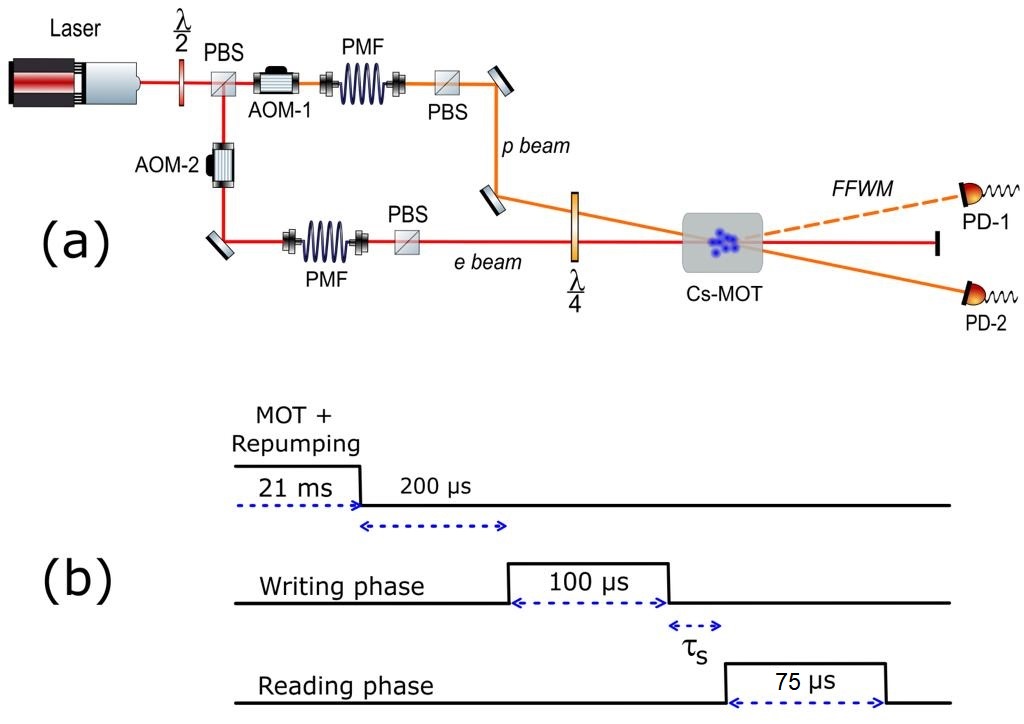}}
  \caption{(Color online) (a): Simplified experimental beams configuration to observe the RIR and the FFWM signals. (b): Time sequence specifying the writing, storage and reading phases. AOM: acousto-optic modulator; PMF: polarization maintaining fiber; PBS: polarizing beam splitter; $\lambda/2$: half waveplate; $\lambda/4$: quarter waveplate; PD: photodetector. }
  \label{fig:Fig1}
\end{figure}

In order to test the validity of the previously developed theory, we have also investigated experimentally the probe transmission and FFWM spectra, both during the writing and reading phases and in transient and  close to the steady state regimes. We have used a cloud of cold cesium atoms with a temperature of hundreds of $\mathrm{\mu K}$ and an on-resonance optical density of about $3$, obtained from a magneto-optical trap (MOT). The experiment is performed in the absence of the trapping and repumping beams as well as the MOT quadrupole magnetic field, with the trapping beams and the quadrupole magnetic field being switched off $1 \,\mathrm{ms}$ before the repumping beam in order to pump the atoms in the hyperfine $6S_{1/2} (F=4)$ ground state. The residual magnetic field is compensated by three pairs of independent Helmholtz coils, whose current is adjusted by using a microwave spectroscopic technique \cite{Allan16}, which allows the residual field to be reduced to less than $10 \,\mathrm{mG}$. The simplified experimental scheme is shown in Fig. \ref{fig:Fig1}(a). All the beams are provided by an external-cavity diode laser locked using saturated absorption signal to the cesium closed transition  $6S_{1/2} (F=4) \rightarrow  6P_{3/2} (F^\prime =5) $ . The excitation beam (E) and the probe beam (P) have the same circular polarization and their directions form a small angle of $\theta=1\degree$. The amplitude and frequency of beams E and P are controlled by independent acousto-optic modulators (AOMs). The frequency of the excitation beam is red-detuned by $10 \,\mathrm{MHz}$ from the transition  $(F=4) \rightarrow(F^\prime =5)$, while the frequency of the probe beam is scanned around the frequency of the excitation beam. The probe transmission and the generated FFWM intensities are detected by fast photodetectors. The time sequence of the experiment is shown in Fig. \ref{fig:Fig1}(b). The excitation and probe beams are kept on for a period of $100\,\mathrm{\mu s}$ during the writing phase and then they are both turned off for a controlled period of time, $\tau_S = t_2 - t_1$. After this storage phase, the excitation beam E is turned back on to retrieve both the probe and  the FFWM beams, which are detected by the same pair of photodetectors. 

\begin{figure}[!htb]
\includegraphics[scale=0.3]{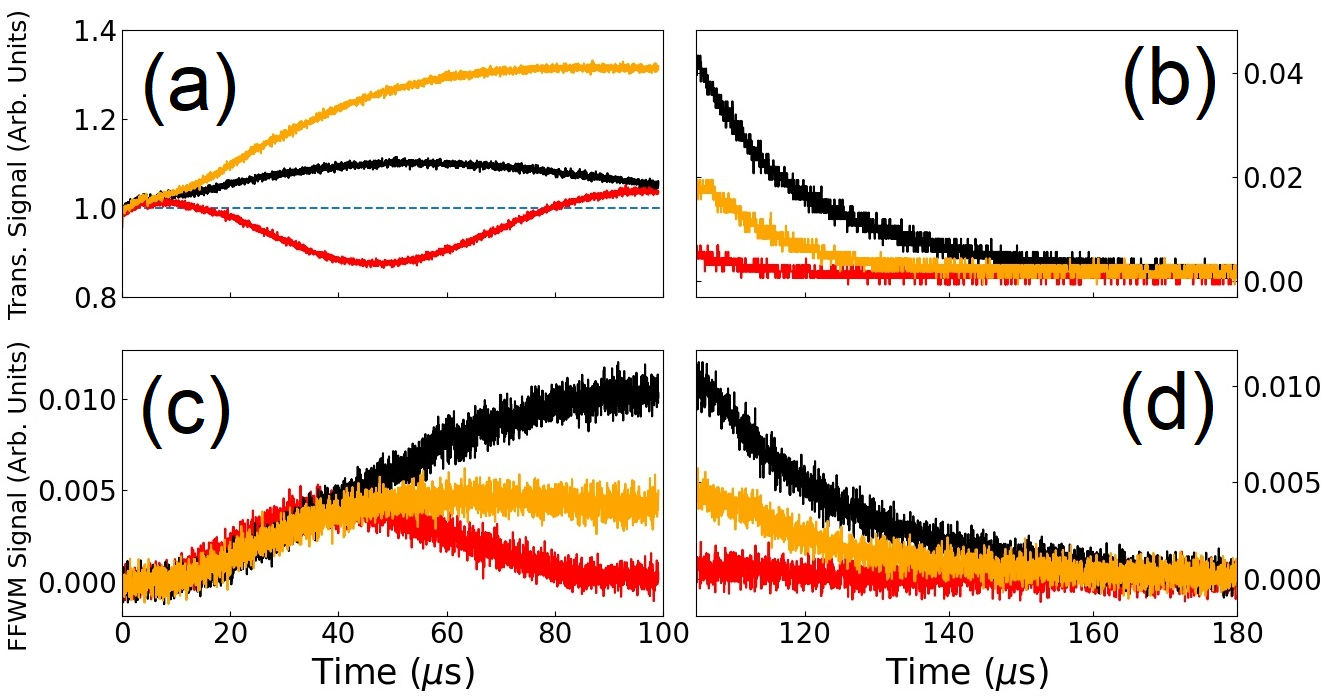} 
 \caption{(Color online) Time evolution of the probe transmission and the generated FFWM beams during the writing, (a) and (c), and the reading, (b) and (d), phases for $\delta \approx -8 \,\mathrm{kHz}$ (orange/light gray), $\delta \approx 0\,\mathrm{kHz}$ (black) and $\delta \approx +8 \,\mathrm{kHz}$ (red/dark gray). In (a) we have normalized the signal associated by the intensity of the incident probe, as explained in the text.}
  \label{fig:Figsignals}
\end{figure}

In the experiment, for a given excitation-probe frequency detuning $\delta$, we record the signals propagating along the directions of the incident probe beam and the generated FFWM beam, both in the writing and reading phases. In Fig. \ref{fig:Figsignals} we show these signals as a function of time for excitation-probe detunings of $\delta \approx -8\, \mathrm{kHz}$, $\delta \approx 0\, \mathrm{kHz}$, and $\delta \approx +8\, \mathrm{kHz}$, respectively. In panels (a) and (c), we plot the curves for the writing period of $100 \,\mathrm{\mu s}$. In panels (b) and (d), the time evolutions are plotted for a reading period of $75\,\mathrm{\mu s}$ after a storage period of $\tau_S = 5\,\mathrm{\mu s}$. The power of the excitation and probe beams are equal to $50\,\mathrm{\mu W}$ and $1\,\mathrm{\mu W}$, respectively. First, we should note that the probe transmission signal is indeed associated with homodyne detection between the generated signal propagating along the probe beam and the incident probe beam, which correspond to the measurement of the imaginary part of the nonlinear susceptibility, while the FFWM signal is a measure of its squared modulus. Therefore, in the frame (a) of Fig. 8 we have normalized the measured probe transmission signal by the incident probe intensity. The retrieval of the probe and FFWM signals in the reading phase demonstrates the information on these beams have been stored in the atomic ensemble. As can be observed in panels (a) and (c), both the probe transmission and the generated FFWM signals present a transient regime strongly dependent of the excitation-probe detuning. This effect have already been observed on previous RIR experiments for the transmitted probe~\cite{Guibal96}. Our results show the same transient regime is also present in the generated FFWM signal. Nevertheless, all the signals reach a stationary regime even in the absence of any homogeneous decay rate as predicted by the developed theoretical model. It is also worth noting that both the retrieved signals, at the beginning of the reading phase, are always maximum for zero detuning, with their decays coming from the atomic motion, as we have already reported in \cite{Allan16}. The amplitudes measured for the retrieved signals along the probe and FFWM directions have the same order of magnitude, but the signal in the direction of the probe beam is still larger by a factor of four. Our theoretical analysis predicts the same amplitudes for the retrieved signals, and presently we do not fully understand this discrepancy. One possible cause could be a larger sensitivity of the FFWM signal to the phase matching conditions in the system, with beams of finite transversal dimensions.

\begin{figure}[!h]
\includegraphics[scale=0.32]{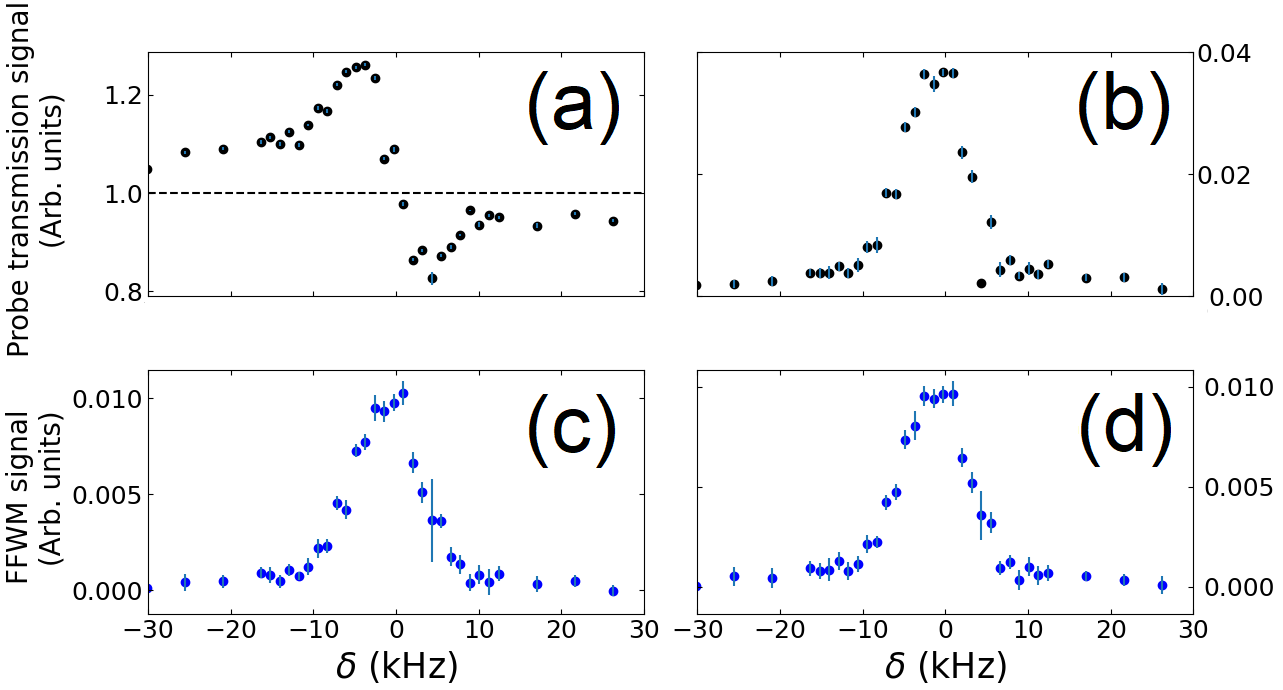} 
\caption{Measured spectra for the probe transmission in (a) the end of the writing phase and (b) the beginning of the reading phase, and corresponding spectra for the FFWM in the (c) writing and (d) reading phases.}
\label{fig:Fig7}
\end{figure}

Figure \ref{fig:Fig7} shows the measured signal intensity spectra. The intensities are measured at the end of the writing phase and at the beginning of the reading phase. We clearly see that the retrieved spectra for the signal propagating along the directions of the probe and the FFWM beams have essentially the same spectral width as the corresponding signals in the writing phase, which evidentiates they are determined by the same physical mechanism. We may also note the similar structure of the spectra predicted by the theoretical model introduced in section II.

\section{Discussions} \label{discussions}

\begin{figure}[!htb]
  \vspace{0.1cm}
	\includegraphics[scale=0.3]{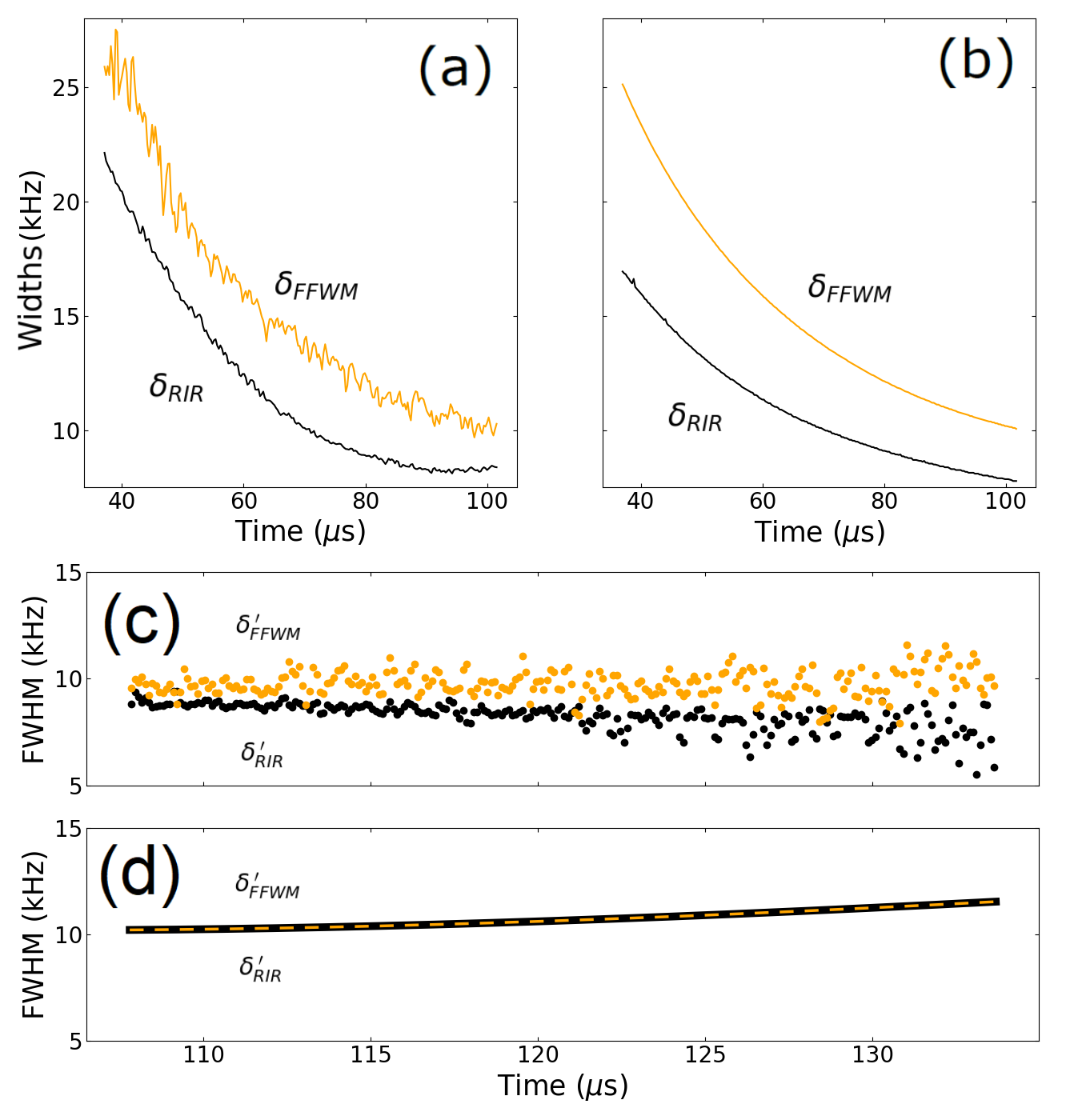}
  \caption{(Color online) (a): Experimental temporal evolution of the spectral width of the probe transmission spectrum (frequency separation between the gain peak and the absorption peak)  and the full width half maximum of the generated FFWM spectrum measured in the writing phase. (b): Theoretical curves corresponding to (a).  (c): Experimental temporal evolution of the spectral full width half maximum of the retrieved FFWM and transmission spectrum measured in the reading phase. (d): Theoretical curves corresponding to (c).}
  \label{fig:Fig9}
\end{figure}

In this section, we aim to provide at first a global comparison between the theoretical model predictions and the experimental results, secondly we study the time profiles of these signals and delve into the limits of our theoretical model. During the writing phase of the experiment, we expect a progressive establishment of coherence between different momentum states which translate into a transient dynamics of the measured spectra that converges to a stationary shape. During the reading phase, the loss of coherence due to atomic motion implies that the shape of the measured spectra is only slightly time-dependent. In order to achieve the first goal of this section, we obtain the time evolution of the linewidths of each signal (for the writing and reading phases). The signal's linewidths (for the probe transmission signal the linewidth is defined as the frequency separation between the gain and absorption peaks) in the writing phase are strongly dependent on the time we perform the measurement. For very short times, the widths are large and Fourier limited, evolving to stationary values for long times. This behavior was verified experimentally as shown in Fig \ref{fig:Fig9}(a) and is in reasonable agreement with the predictions of the developed theory as shown in Fig. \ref{fig:Fig9}(b). We note that the construction of coherence makes the measured spectra sharper in time. As it was demonstrated previously in \cite{Grynberg94} the stationary value of the probe transmission signal linewidth is determined by the temperature of the atomic ensemble. Therefore, our results suggest a new way to measure the temperature of the atoms through the measurement of the linewidth of the FFWM signal, which is background free. Figures \ref{fig:Fig9}(c) and \ref{fig:Fig9}(d) show the experimental and theoretical time evolution for the retrieved linewidths, which theoretically get larger in time and one should note that an experimental observation of this dynamics becomes very limited by noise for longer times. Note that despite the good global agreement, the theoretical model does not show any difference between the probe and FFWM linewidths in the reading phase, even though the experiment shows a slight difference of about $1\,\mathrm{kHz}$.\\

\begin{figure}[!htb]
\centering
\includegraphics[scale=0.25]{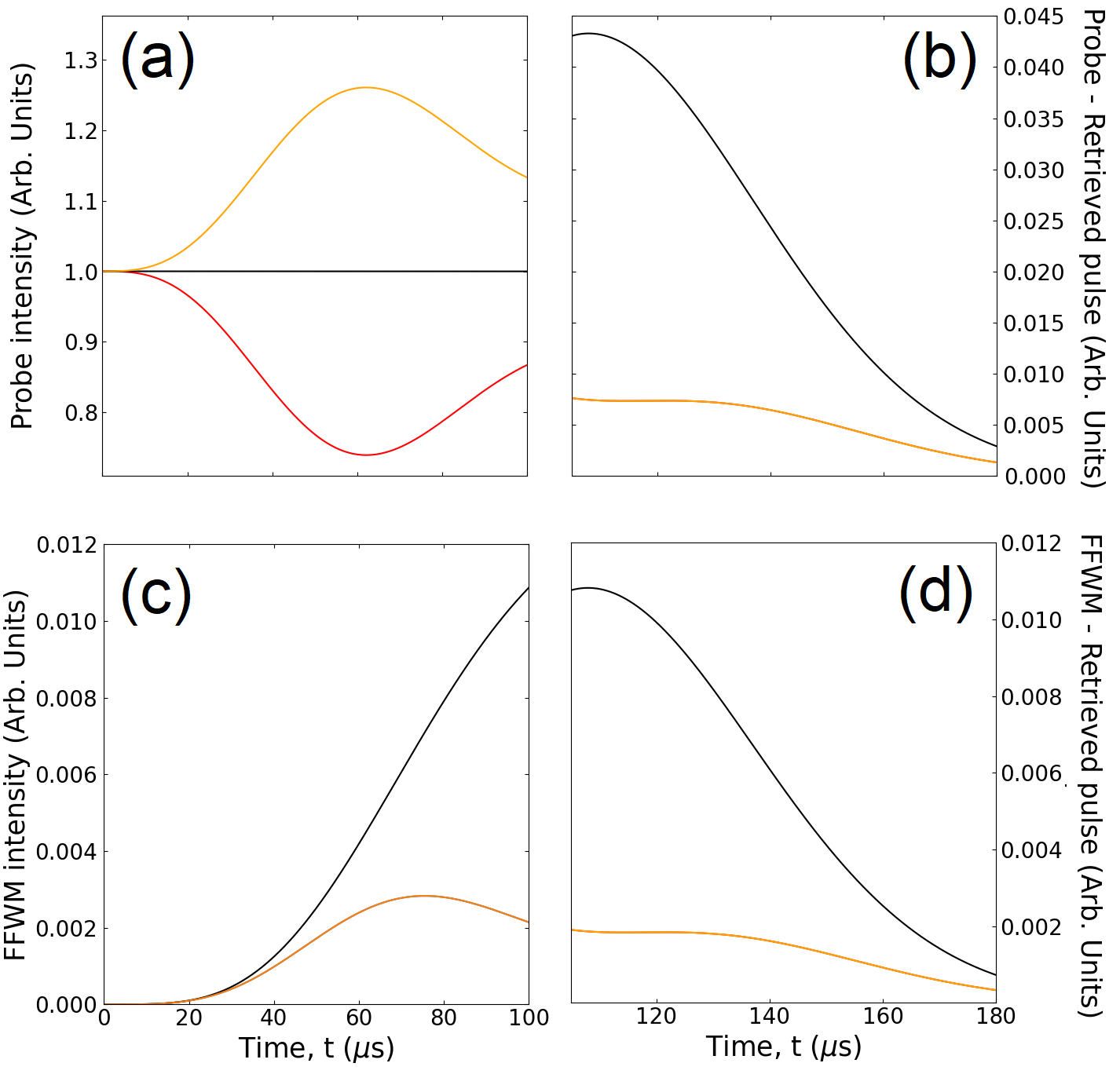} 
 \caption{(Color online) Theoretical time evolution of the probe transmission signal in the (a) writing and (b) reading phases. Generated FFWM signals during the (c) writing and (d) reading phases for: $\delta \approx -8 \,\mathrm{kHz}$ (orange/light gray), $\delta \approx 0\,\mathrm{kHz}$ (black) and $\delta \approx 8 \,\mathrm{kHz}$ (red/dark gray).}
  \label{fig:profiles}
\end{figure}

A closer look at the theoretical time profiles of the signals provided in Fig. \ref{fig:profiles} shows some limitations of our model. In fact, comparing Figs. \ref{fig:Figsignals}(a) and \ref{fig:Figsignals}(c) to Figs. \ref{fig:profiles}(a) and \ref{fig:profiles}(c), for example, we note that the experimental signals are noticeably more asymmetric than the experimental ones. More importantly, Figures \ref{fig:Figsignals}(b) and \ref{fig:Figsignals}(d) show experimental decay times lower than the ones predicted by our theory, and the shape of the experimental signal follows a more exponential-like decay than a Gaussian-like one, as predicted by our theoretical model. In Ref.~\cite{Allan16}, however, we had previously reported a curve corresponding to Fig.~\ref{fig:Figsignals}(b) that followed more a Gaussian-like shape. In our view, this discrepancy may result from different structures of the residual magnetic field in our experimental apparatus. For example, if the residual field contains a larger component perpendicular to the propagation direction, it may remove atoms from the cycling transition we pumped the atoms into. We could then introduce this mechanism as a simple decaying exponential depletion of atoms that participate in the process. In this manuscript, however, we opted not to introduce any phenomenological modification of the theory, in order to highlight the limits of our first-principles approach. Additional effects such as induced heating of the atomic ensemble by the interacting fields are also not included in our theoretical model but could play a role in the time dynamics, as already pointed out on \cite{Allan16}.

\section{Conclusions}
In this paper we presented a theoretical model from first principles for a RIR based atomic memory and provided experimental results to corroborate it. Our theory puts the probe transmitted signal in equal footing to the FFWM signal. This is first directly illustrated via the calculations for the of the temporal evolutions and lineshapes during the reading process, which showed the same structure and amplitude for both signals. Furthermore, the theory predicted that both probe transmission and FFWM are connected to non-volatile memories, as previously observed only for the transmitted signal. This means that the stored in information can be retrieved without its simultaneous destruction. These signals were then in fact experimentally observed with good agreement with the core theoretical predictions. For the writing and reading phase, both (probe and FFWM) experimental spectra showed the same structures predicted by the model. The global comparison between theory and experiment was then carried out in more detail through the time evolution of the linewidths experimentally observed and theoretically predicted for the structures in the writing and reading phases. This comparison showed a good qualitative agreement between experimental data and theoretical predictions, even though some differences could still be noticed. The fact that the theory shows the FFWM and transmission signals as originating from the same phenomenon, at the same order of perturbation, suggests the use of this pair of signals as a source of quantum correlations. Recently, our group reported the observation of the analogous to the non-volatile memory for the spontaneous scattering of light from an ensemble of two-level atoms at the single photon level~\cite{Moreira2021}. We also reported the observation of non-classical correlations in the continuous generation of photon pairs in the backward four-wave-mixing excitation of an ensemble of two-level atoms~\cite{Araujo2021}, an effect theoretically proposed in 2007~\cite{Du2007}. Thus, our present work point out to the possibility of extending these previous results to explore the correlations in the probe transmission and FFWM at the single photon level with memory.

\vspace{0.5cm}
\noindent
\begin{center}
{\bf \large Acknowledgments}
\end{center}

This work was supported by the Brazilian funding agencies CNPq (program INCT-IQ, No. 465469/2014-0), CAPES (program PROEX 534/2018, No. 23038.003382/2018-39), and FACEPE (program PRONEM 08/2014, No. APQ-1178- 1.05/14). It was also funded by the Public Call n. 03 Produtividade em Pesquisa PROPESQ/PRPG/UFPB proposal code PVA13253-2020.

\end{document}